\documentclass[10pt]{article}

\usepackage{amsmath}
\usepackage{amssymb}

\usepackage{graphicx}

\usepackage{cite}

\usepackage{hyperref}

\usepackage{lineno}

\usepackage{microtype}
\DisableLigatures[f]{encoding = *, family = * }

\usepackage{setspace} 
\usepackage[labelfont=bf,labelsep=period,justification=raggedright]{caption}

\makeatletter
\renewcommand{\@biblabel}[1]{\quad#1.}
\makeatother

\date{}

\begin{document}

\begin{flushleft}
{\Large
\textbf{Unpacking of a crumpled wire from two-dimensional cavities}
}
\\
Thiago A. Sobral$^{1}$, 
Marcelo A. F. Gomes$^{1\ast}$,
N\'ubia R. Machado$^{2}$,
Valdemiro P. Brito$^{2}$
\\
\bf{1} Departamento de F\'isica, Universidade Federal de Pernambuco, 50670-901, Recife, PE, Brasil
\\
\bf{2}  Departamento de F\'isica, Universidade Federal do Piau\'i, 64049-550, Teresina, PI, Brasil
\\
$\ast$ E-mail: mafg@ufpe.br
\end{flushleft}


\section*{Abstract}

The physics of tightly packed structures of a wire and other threadlike materials confined in cavities has been explored in recent years in connection with crumpled systems and a number of topics ranging from applications to DNA packing in viral capsids and surgical interventions with catheter to analogies with the electron gas at finite temperature and with theories of two-dimensional quantum gravity. When a long piece of wire is injected into two-dimensional cavities, it bends and originates in the jammed limit a series of closed structures that we call loops. In this work we study the extraction of a crumpled tightly packed wire from a circular cavity aiming to remove loops individually. The size of each removed loop, the maximum value of the force needed to unpack each loop, and the total length of the extracted wire were measured and related to an exponential growth and a mean field model consistent with the literature of crumpled wires. Scaling laws for this process are reported and the relationship between the processes of packing and unpacking of wire is commented upon. 
\newpage


\section*{Introduction}

The process of extraction or unpacking of an object which has an effective one-dimensional topology is very common in nature, including the DNA molecule that is ejected from virus capsids~\cite{Purohit05} and the extraction of catheters in surgical interventions~\cite{Catheter1,Catheter2} up to the unwinding of wires in industry, and the unpacking of polymers and long-chain biomolecules in drug delivery~\cite{DDS1,DDS2,DDS3,DDS4,DDS5,DDS6,DDS7,DDS8,DDS9}. The initial conformation of such one-dimensional objects confined in a finite volume present some analogies with the patterns found in the packing or in the crumpling of a long piece of wire~\cite{Aguiar91,Vetter14}. The systematic study of crumpling processes began more than two decades ago with crumpled surfaces obtained from sheets of paper and aluminium foils submitted to rapid and ill-defined deformations typical of haphazard procedures aiming to confine them into small volumes~\cite{Kantor86,Gomes87}. The packing of DNA in viral capsids has in recent years been associated with two-dimensional packing of wires, which presents the same fractal dimension~\cite{Maycon08,Katzav06}. 

The research of crumpled wires in two dimensions, on the other hand, began in the last decade~\cite{Donato02,Donato03}. When one injects, for instance, a long wire of copper into a two-dimensional cavity that allows only a single layer of wire, inner structures of iterated loops are formed with a pattern of points of contacts generating jammed structures as those observed the classical problem of packing of discs~\cite{Sloane98}. The general two-dimensional pattern associated with the packing of a long piece of wire presents some morphological phases~\cite{Boue06,Boue07,Stoop08} which are related with physical properties of the structure. Recently, crumpled wires have been studied in connection with striking analogies with other different systems such as the electron gas at finite temperatures~\cite{Gomes10,Gomes13}, and two-dimensional quantum gravity~\cite{Cunha09}. 

In the present paper we are interested in the {\it unpacking} of a crumpled wire from two-dimensional cavities. In this case, when the wire is extracted with the aid of a dynamometer it is possible to record the force involved in the extraction and relate it with both the size of the loop which is released and the extracted length of the wire. A profile for the force is of great interest as indicated in several previous studies~\cite{Boue06,Boue07,Stoop08,Cunha09,Bayart14}, but we did not find in the literature any paper concerning processes of unpacking of wires from two-dimensional cavities. In spite of the introduction of some level of irreversibility in the packing of wire in a cavity as a consequence of high values of deformation and due to the effects of plasticity ever present, important results are achieved. Basically we found that the force needed to extract the wire has a power law dependence with the size of the loops which it is better explained by the global features of the system. We also observed that the size of the loops decreases exponentially with its order of unpacking, allowing us to propose a differential equation for the (un)packed system. The overall framework is consistent with previous studies where the energy involved in the packing processes is well described by models of ordered folding~\cite{Deboeuf13}.

\section*{Experimental details}
\begin{figure}[!b]
\centering
\includegraphics[width=0.6\linewidth]{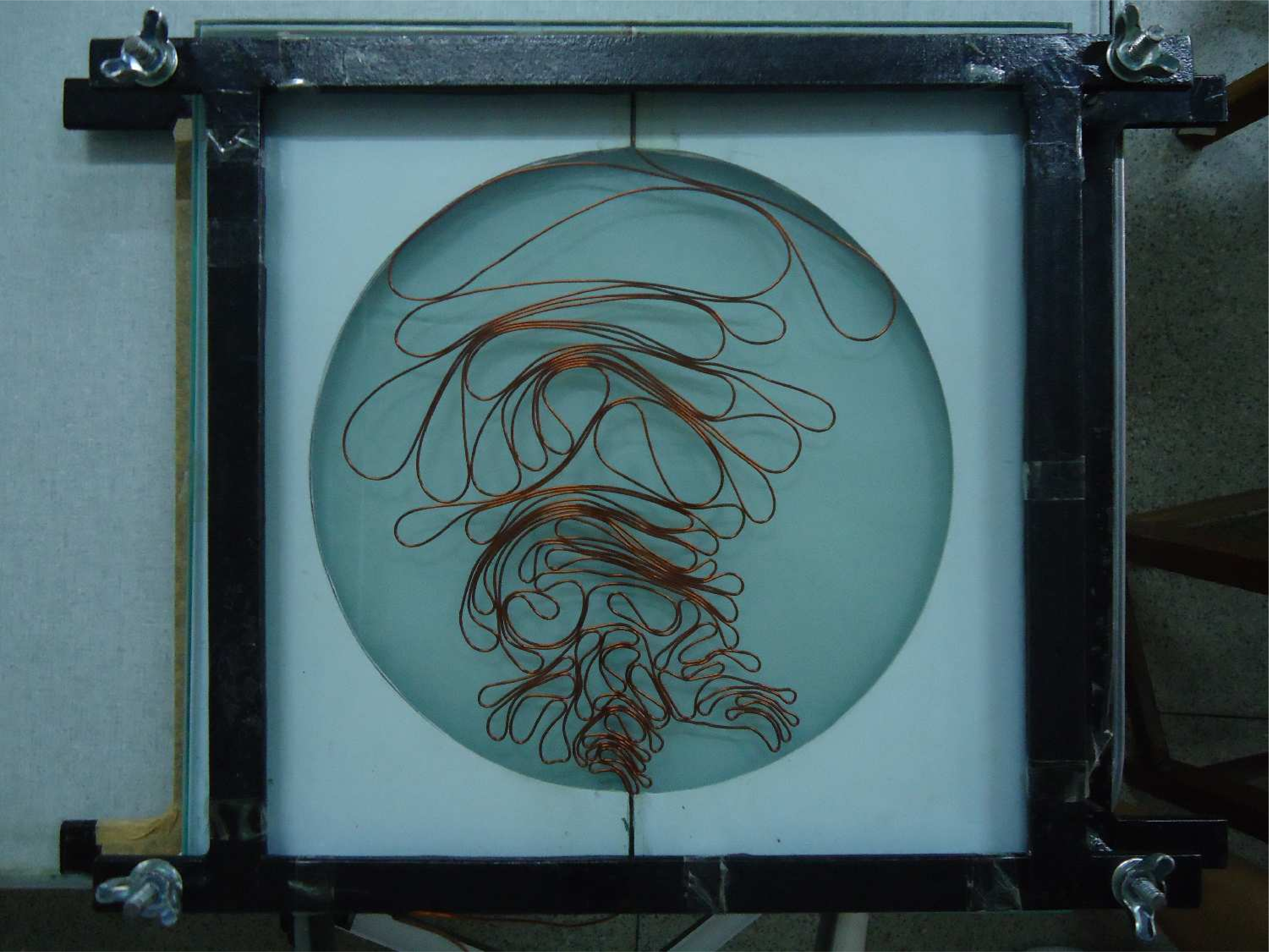}
\caption{{\bf A wire that is 5020 mm long at the jammed state in a circular cavity whose diameter is 22 cm.} Our experiment consists in measuring the force needed to pull that piece of copper outside the cavity.}
\label{apparatus}
\end{figure}

The cavities used in our experiments consist in two plates of glass of 300 mm x 300 mm x 8 mm separated by  circular acrylic molds of diameter $d=$ 15 cm and $d=$ 22 cm and 1.1 mm height that allows the wire to be accommodated  without superposition. The simply connected cavities used in our experiments have two channels at opposite sides of the molds for the injection of the wire, but in this study one of them is used to fix one of the wire ends (Fig.~\ref{apparatus}). 
The wires used were made of copper and had a diameter of $\zeta = 1$ mm. Initially they were pushed manually into the cavity through one of the channels with an approximately uniform speed of about 1 cm/s until the system reached the jammed state. The experiments are performed in a dry regime without any lubrication, and its average velocity was estimated by the time needed to inject the wire. The inner structures formed in the injection process, shown in Fig.~ \ref{apparatus}, are quite similar to others previously reported in the literature~\cite{Donato02, Donato03}, and they were used as our initial condition. A heterogeneous cascade of loops (units that have a bulge in one extremity, and two branches of the wire merging in the other extremity) is distributed by the available inner area of the cavity so that the smallest loop is adjacent to the injection channel, while the largest one tends to be located as distant as possible from this channel. The rigid crumpled structures formed in the jammed limit presented on average a maximum packing fraction of $0.16$ for both cavities.

For each diameter of the cavity, we repeated the following procedure for eight equivalent experiments: ({\it i}) after the jamming state is reached, we initiate the unpacking process; ({\it ii}) a digital dynamometer is attached to the outer end of the wire such that the force needed to unmake a single loop can be measured within $1\%$ uncertainty and recorded; ({\it iii}) the wire is marked with a permanent marker at the point where the loop starts to be unmade; ({\it iv}) after extracting more and more wire the loop is totally unmade and the point associated with the second extremity of the loop is also marked. The difference between two consecutive marks is identified as the size of that particular loop. It is clear from the experimental procedure that the first (smaller) extracted loops are harder to pull than the successive (bigger) ones which is compatible with the intuitive idea of a jammed state. 

\begin{figure}[!b]
\centering
\includegraphics[width=0.6\linewidth]{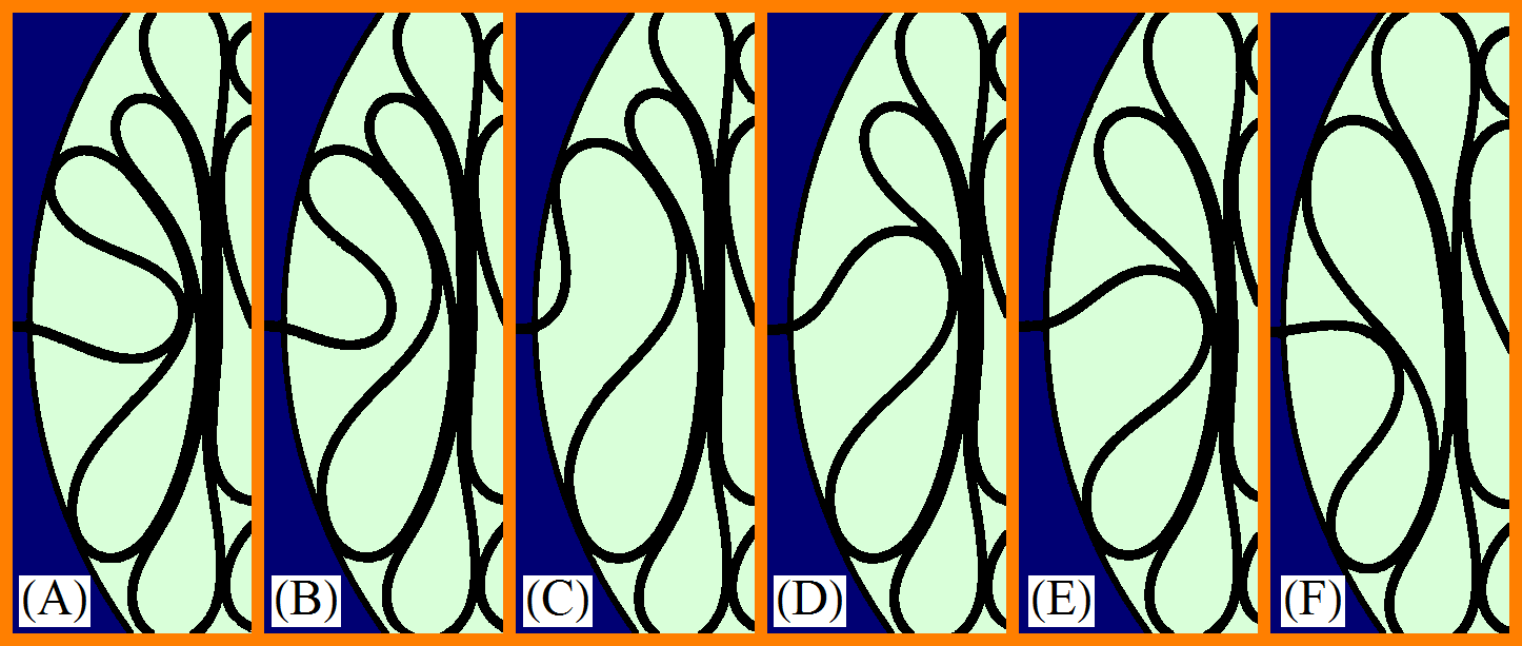}
\caption{\bf Sequence of images in a real experiment showing the extraction of a single loop from a two-dimensional cavity.}
\label{fig2}
\end{figure}

Fig.~\ref{fig2} (from A to F) is a sequence of images of the unpacking of a single loop from a circular cavity, where the size of the loop is the extracted length between the images A and F.  
It is important to emphasize that, contrarily to what could appear intuitively, the structure of loops within the cavity is self-blocking and it is maintained static even with reduced friction. This result is, in fact, supported by additional experiments in which cavity and wire were previously treated with mineral oil. This happens because the passage through the extraction channel is a very strong constraint for the curvature of the wire. It can be noticed that the elastic structure inside the cavity tends to expand and it increases the difficulty of the alignment of the wire beyond the full extension of the extraction channel. During the extraction of a loop there is a restoring force, but after each extraction the system finds a new equilibrium state. Here it was useful as an advantage for the usage of the permanent marker. The profile of the force as a function of the length presents its maximum magnitude near the end of the extraction of the loop (Fig.~\ref{fig2}D and \ref{fig2}E). Roughly speaking, the functional shape of the force is repeated for each extracted loop, with a progressive decrease of its maximum magnitude, as the size of the loops continues to increase.

It is interesting to note that the reverse sequence (from F to A) in Fig.~ \ref{fig2} could illustrate the insertion of the loop as well. This last observation have inspired us in several points of this study, as it can be seen in the following pages. An important difference between the present work and the previous studies of packing of wire is that here the total number of loops and the total length inside the cavity are both initially known quantities.

\section*{Results and Discussions}

\subsection*{The size of a loop}

\begin{figure}[!ht]
\centering
\includegraphics[width=0.6\linewidth]{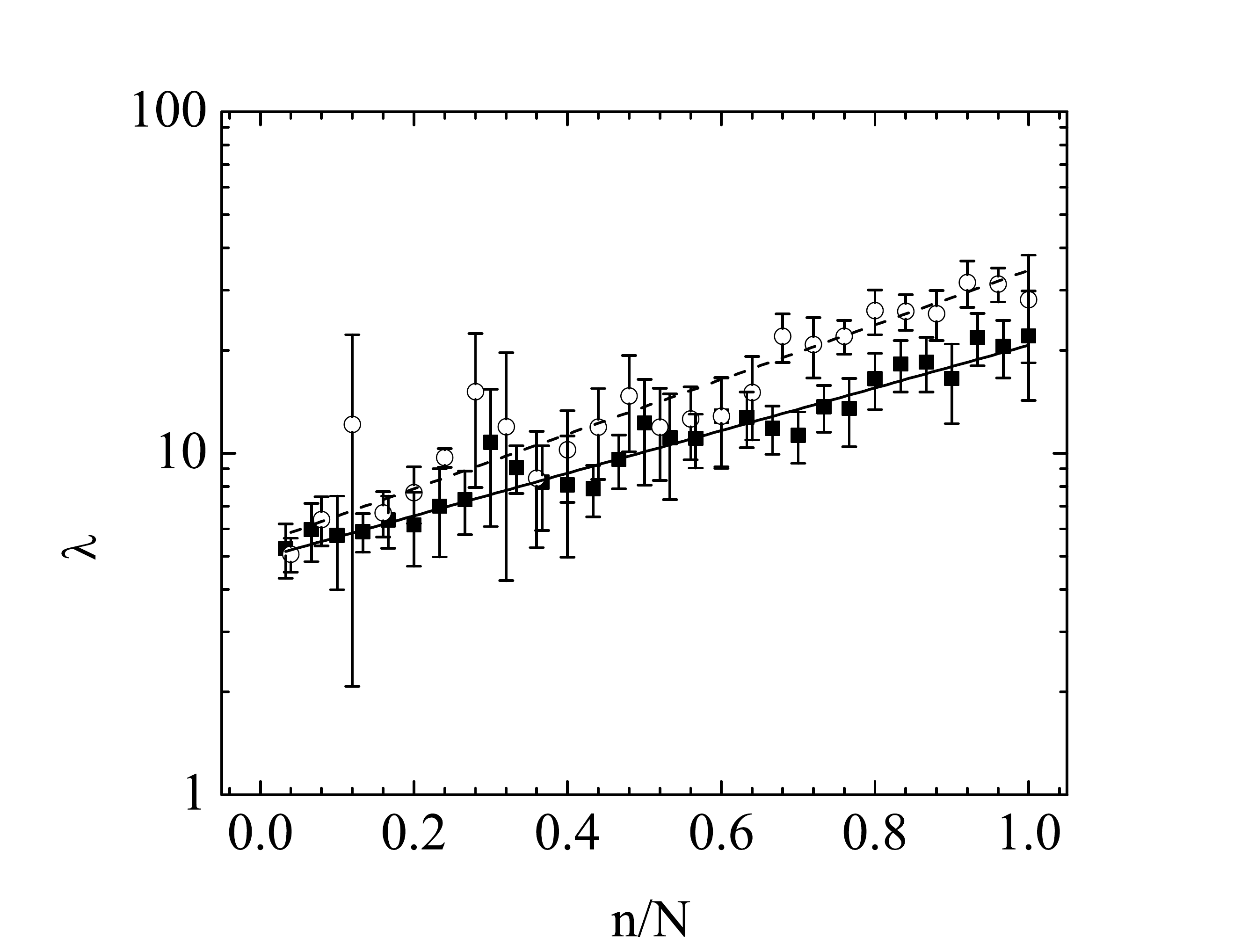}
\caption{{\bf The average size of loops (i.e. arc length), $\lambda$ (cm), as a function of the fraction of its unpacking order, $n/N$, for copper wires in cavities with $d  = 15$ cm (black squares) and $d  = 22$ cm (opened circles).} The lines show the best fit of Eq.~\ref{eq1} for the data with $d  = 15$ cm (thick continuum line) and $d  = 22$ cm (dashed line). See text for details.}
\label{fig3}
\end{figure}

Every experiment of jammed state $i$ had its total number of loops $N_i$ recorded. However, even under similar conditions, the total number of loops $N_i$ varies among the experiments. In our experiments we obtained $N_i$ ranging from 24 to 42, for $d$ = 15 cm, and ranging from 24 to 38, for $d$ = 22 cm. In order to have an universal domain to take averages, the number of loops remaining in the cavity $n_i$ were normalized by $N_i$, so that $n_i/N_i$ becomes a number that runs from zero to one. In order to assure that the average total number of loops, $N$, is the final quantity of points, our averages are taken over intervals of $N^{-1}$ of width, over all data. Follows from this procedure that each point represents, in average, eight experimental data. The result for the average size (i.e. arc length) of the loop, $\lambda$, as a function of the fractional order of unpacked loop, $n/N$, is shown in Fig.~\ref{fig3} for copper wires in cavities of 15 cm and 22 cm of diameter.

From Fig.~\ref{fig3}, it can be seen that the size of the smallest loop have the same values, within the error bars, for $d$  = 15 cm and for $d$  = 22 cm. This suggests that the minimum value for the size of the loop does not depend on the diameter of the cavity and, instead, depends only on features of the wire (the plasticity imposes a cutoff in size). On the other hand, the last unpacked loop seems to be proportional to the diameter of the cavity~\cite{Donato02,Gomes11}. The standard error of the data shows the high fluctuations involved in a typical experiment of packing of wire~\cite{Gomes11}. The overall growth of the data in Fig.~\ref{fig3} resembles an exponential behavior 
\begin{equation}
\lambda_n  = \lambda_0 e^{R_0 (n/N)},
\label{eq1}
\end{equation}
where $\lambda_0$ is an initial size obtained by the $n/N \rightarrow 0$ extrapolation, and $R_0$ is a continuum growth rate. Fits from the experimental data with Eq.~\ref{eq1}  are represented in Fig.~\ref{fig3} by a full line for $d=$ 15~cm and by a dashed line for $d=$ 22~cm. The parameters found were $\lambda_0^{(15)} = 4.9 \pm 0.2$ cm, and $R_0^{(15)} = 1.44 \pm 0.06$ for $d$  = 15~cm, and $\lambda_0^{(22)} = 5.4 \pm 0.2$~cm, and $R_0^{(22)}  = 1.84 \pm 0.07$ for $d$  = 22~cm. As commented before, the expected physical invariants are $\lambda_0$, the smallest scale length, and $\lambda_N$, the scale length ruled by the cavity. Consequently, $R_0 = \ln(\lambda_N / \lambda_0)$ does not vary among experiments performed in a same cavity (unlike $N_i$). The parameters found correspond to $\lambda^{(15)}_N = (21 \pm 2$)~cm and $\lambda^{(22)}_N = (34 \pm 3$)~cm for the size of the last unpacked loop. The total growth of the loops is expected to scale with the diameter of the cavity, $\lambda_N / \lambda_0 \sim d$, which implies that $\Delta R_0 = \Delta (\ln d)$; here $\Delta R_0 = R_0^{(22)}  - R_0^{(15)}  = 0.40$ while $\Delta (\ln d) = \ln(22/15) = 0.38$, a small deviation of about $5\%$. This suggests that the present model is consistent among different cavities.

An exponential growth as stated in Eq.~\ref{eq1} is obtained from the hypothesis that each loop occupies a fraction of the available space of the cavity discounting early loops. However, in the literature of packing of wires in two-dimensional cavities there is a hierarchical model~\cite{Donato02} that fits the asymptotic limit $n/N \rightarrow 0$. That model is based on fractal scaling (iterations) and provides a more complicated relation with a bigger number of parameters. In the present study we pulled the wire loop-by-loop providing a very detailed profile of $\lambda(n)$ and observed that Eq.~\ref{eq1} fits sufficiently well the data obtained (Fig.~\ref{fig3}) and involves a better insight of the physics of the problem. For these reasons we adopt an exponential model here.

\subsection*{The force to unpack a single loop}

\begin{figure}[!ht]
\centering
\includegraphics[width=0.9\linewidth]{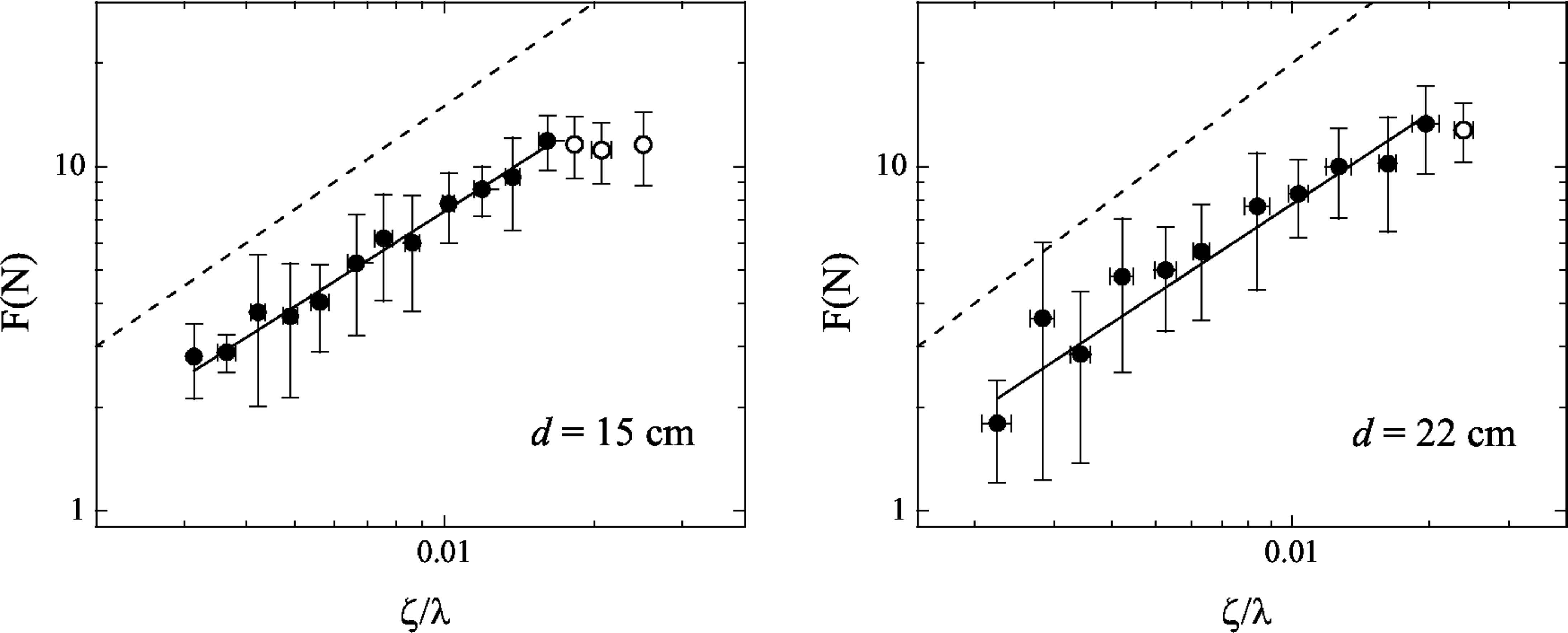}
\caption{{\bf The force $F$ in Newtons needed to pull a loop out the cavity as a function of $\zeta / \lambda$, for copper wires in cavities of diameter $d=15$~cm and $d=22$~cm.} The dashed lines are guides for the eyes and represent the linear regime. See text for details.}
\label{fig4}
\end{figure}

In order to examine the force $F_j$ needed to unpack a loop of size $\lambda_j$, the data were binned over intervals of $\Delta \lambda$ and then an average of the force, $F$, was taken as well as an average of the size, $\lambda$. Fig.~\ref{fig4} (left) shows the result of that procedure for the data of the cavity with $d  = 15$ cm. A power law
\begin{equation}
F \sim  \left( \zeta / \lambda \right)^{\alpha}
\label{eq2}
\end{equation}
fits the data very well except for high values of $\zeta / \lambda$ (small values of $\lambda$). The $\alpha$ value found for the fit shown in the Fig.~\ref{fig4} (left) was $\alpha^{(15)} = 0.92 \pm 0.03$. Fig.~\ref{fig4} (right) shows the data for the cavity with $d  = 22$ cm, and the power law best fit found had $\alpha^{(22)} = 0.87 \pm 0.05$. We ignore the three last points in Fig.~\ref{fig4} (left) as well the last point in Fig.~\ref{fig4} (right) because they deviate from the general tendency since the elastic behavior fails and plasticity takes place. These points are represented by open circles.

We observe that the available data shown in Fig.~\ref{fig4} have a narrow range for the size of the loops. This is a very important point that we can not overcome right now. Following the relations $R_0 = \ln(\lambda_N / \lambda_0)$ and $\Delta R_0 = \Delta \ln(d)$, we estimate that one decade in $\lambda$ is achieved for cavities of about $d=$ 35 cm. Our wire has a diameter $\zeta = 1.0$~mm. This entails a corresponding diameter ratio ($d/\zeta$) of at least 350, a number that impose serious practical problems to be circumvented in the machining process of the cavities, especially for transparent cavities (glass or acrylic), as in our case. The diameter ratios in our experiment are 150 and 220, and the maximum diameter ratio in the literature is about 250 \cite{Stoop08,Bayart14}. From this we believe that, although with a limited range, our results have valid contributions for all current studies of crumpled wires in two-dimensions.

\begin{figure}[!h]
\centering
\includegraphics[width=0.4\linewidth]{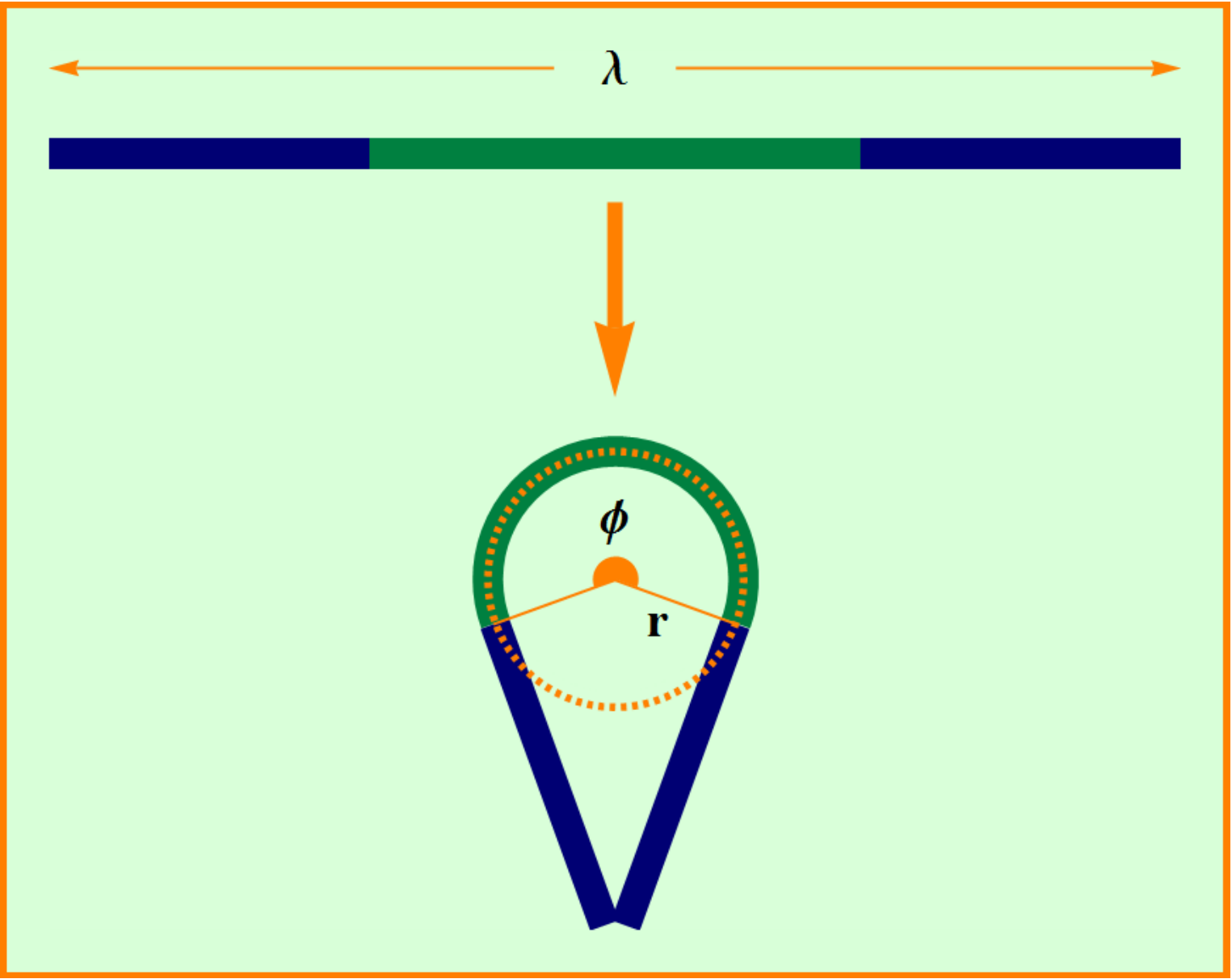}
\caption{\bf Scheme of a bend of a straight wire into a loop.}
\label{fig5}
\end{figure}

Fig.~\ref{fig5} illustrates the bending of a straight wire of length $L=\lambda$ to form a loop with a circular bulge with radius of curvature $r$. The bent rod is stretched in the external edge of the bulge of the loop and it is shrunk in its internal edge. The elastic bending energy depends on the curvature $\kappa(l)$ as  
\begin{equation}
E_{b} = I \int_0^L \kappa(l)^2 dl,
\label{eq3}
\end{equation}
where $I$ depends on the (supposedly fixed) cross section of the wire. Eq.~\ref{eq3} shows that any fixed shape will have $E_b \sim \lambda^{-1}$ and therefore $F_b = - \frac{d E_b}{d \lambda} \sim \lambda^{-2}$. This result has the form of Eq.~\ref{eq2} with $\alpha = 2$ and represents the purely elastic bending~\cite{Stoop08,Purohit05}. In this sense the exponents found experimentally are quite anomalous. However, our experimental result is safer since the two exponents found by fitting the data are the same within an error interval of $6\%$ and the result is robust over different binnings. From this, we conclude that a crude model of fixed shape as illustrated by Fig.~\ref{fig5} and  Eq.~\ref{eq3} is not suitable in describing our system. Other ingredients as the confining cell, the associative elasticity of all loops and the gradient of the size of the loops, among others, are pivotal for a realistic description of the unpacking of wire in two-dimensional cavities. As we discuss later on this paper, the exponent $\alpha$ found in the experiments can be approximately obtained if we consider the collective behavior of the loops inside the cavity instead of a model of single loop as discussed in the present paragraph.

\subsection*{The extracted length}

The length of the wire extracted from the cavity can be obtained by summing the length due to each removed loop, $L_n  = \Sigma_{j=1}^{n} \lambda_j$. Thus the fractional length is
\begin{equation}
l_n \equiv \frac{L_n}{L_N} = A \left( e^{R_0 (n/N)} - 1 \right)
\label{eq4}
\end{equation}
where $A  =(e^{R_0} - 1)^{-1}$, and $R_0$ is the same as in Eq.~\ref{eq1}. The new variable, $l_n$, as defined above, is the fraction of the pulled wire with respect to the total length, then it must be $l_0 = 0$ at the begin of the unpacking and $l_N  = 1$ at the end. Eq.~\ref{eq4}  provides a curve that fits all data very well, with  the constants $A$  and $R_0$ depending on the combination of the cavity and the material properties, represented by the parameters $\lambda_N$ and $\lambda_0$, respectively. Moreover, the reader can observe that the parameter $A$ is determined by $R_0$, which indicates that Eq.~\ref{eq4} depends on a single parameter.

\begin{figure}[!h]
\centering
\includegraphics[width=0.6\linewidth]{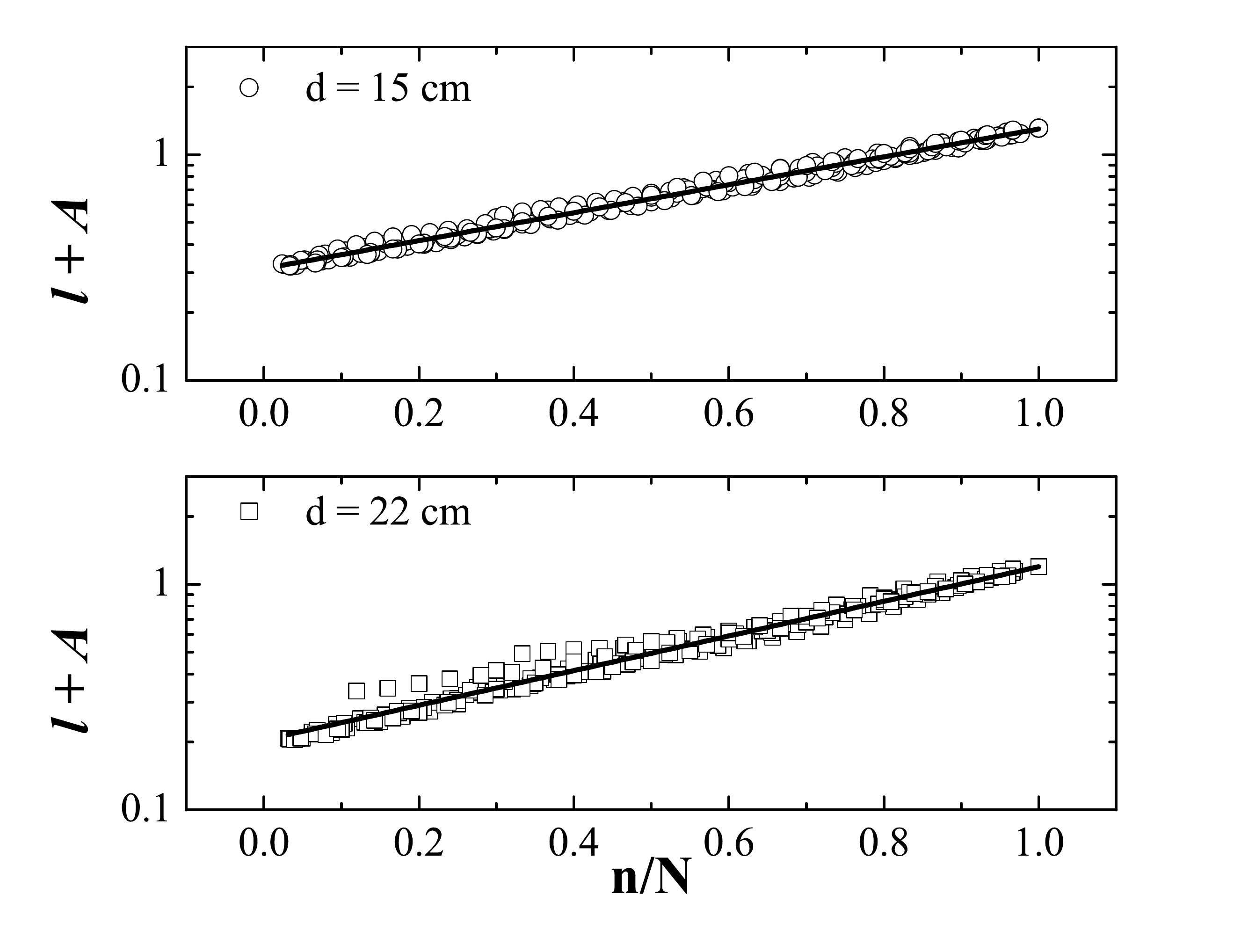}
\caption{{\bf The variable $l+A$ as a function of $n$.} See text for details.}
\label{fig6}
\end{figure} 

Fig.~\ref{fig6} illustrates the best fit of Eq.~\ref{eq4} for the data without binning. For the cavity with $d  = 15$~cm we have found $R_0^{(15)}  = 1.43 \pm 0.05$ and $A^{(15)} = 0.31 \pm 0.02$, while for the cavity of $d = 22$~cm we have found $R_0^{(22)} = 1.78 \pm 0.07$ and $A^{(22)}  = 0.20 \pm 0.02$. These values of $A$ follow its relation with $R_0$. Moreover, these values of $R_0$ are the same found by fitting Eq.~\ref{eq1} to the data from Fig.~\ref{fig3}. They give us $\Delta R_0  = 0.35$, a value $8\%$ distant from $\Delta (\ln d)$.

The exponential model defined by Eq.~\ref{eq1} and Eq.~\ref{eq4} provides 
\begin{equation}
\lambda =  \frac{\lambda_0}{A} \left( l + A \right) ,
\label{eq5}
\end{equation}
which suggests that the size of the loop $\lambda$ is a linear function of the fractional pulled length $l$. It follows from this and Eq.~\ref{eq2} that the force is expected to be a power law of $(l + A)$ with an exponent $\alpha$. This is a second method to obtain the exponent $\alpha$ experimentally, because the high fluctuations in $\lambda$  scrambles the order of the loop in the previous case (observe the data shown in Fig.~\ref{fig2}). A power law obtained from the total length must preserve the order of the unpacking process.

\subsection*{The force to unpack the wire}

\begin{figure}[!h]
\centering
\includegraphics[width=0.9\linewidth]{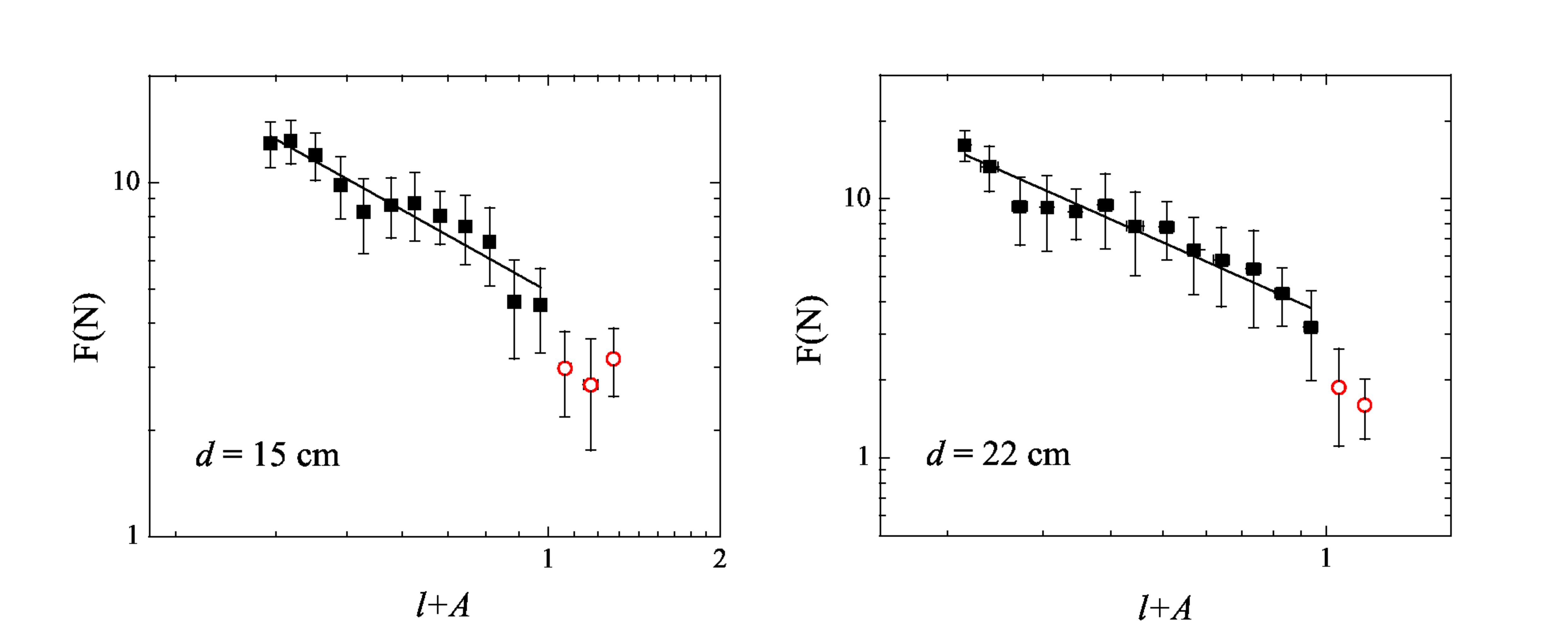}
\caption{{\bf The force $F$ in Newtons needed to pull out a piece of wire whose fraction of length out the cavity is $l$ for copper wires in cavities of diameter $d=15$~cm and $d=22$~cm.} See text for details.}
\label{fig7}
\end{figure}

In order to exhibit the relationship between the force and the length of the wire inside the cavity, an assortment into bins of width $\Delta (l+A)$ is required, where a mean force is taken. Fig.~\ref{fig7} shows the experimental data and the power law best fit for cavities of diameter $d = 15$ cm and $d = 22$ cm, respectively. Both have the form:
\begin{equation}
F  \sim (l+A)^{-\alpha},
\label{eq6}
\end{equation}
and the resulting exponents are $\alpha^{(15)} = 0.90 \pm 0.08$, for $d = 15$ cm, and $\alpha^{(22)} = 0.93 \pm 0.07$, for the cavity of $d = 22$ cm, in the interval of $A \leq (l + A) < 1$. These exponents are equal to those obtained with the direct fits in Fig.~\ref{fig4} within the error bars. However, the exponents increase for $\alpha_2^{(15)} = 1.09 \pm 0.08$ and $\alpha_2^{(22)} =  1.20 \pm 0.08$ if the fits consider all the experimental data. We choose to stop the fits at $l + A = 1$ because after that the data fall from the general tendency as it can be seen by the open circles in Fig.~\ref{fig7}. This happens because the wire is much more loose and loses its contact interactions close to the end of the unpacking.

In the following we give justification for Eq.~\ref{eq6} with $\alpha = 1$. Our model is inspired by previous results from image analysis which indicated that the elastic bending energy in two-dimensional crumpled wires is concentrated in an one-dimensional set, although the mass of the system is distributed in a two-dimensional domain~\cite{Donato07}. This is true for the limit of a fulfilled cavity, where the rod is densely packed as parallel nearly straight lines in the bulk of the cavity and the elastic energy due to the compressed loops is concentrated on the perimeter of the cavity. Interestingly, in this limiting case we can state that {\it the energy is equal for all loops irrespective of its arc length}. Certainly, this conclusion from a particular static configuration does not need to be valid for a serial extraction of loops. However, we argue that in a mean field approach this statement can be extended for the unpacking of wire.

The wire stores elastic energy in the bulge of its loops, which tend to be as big as they can, in a process which involves repulsion among the neighbor loops. This interaction between pairs of loops leads to the exchange of energy among the loops confined in the cavity. Of course, a detailed explanation presents some difficulty because the interaction among the loops is complex as can be inferred from Fig.~\ref{fig2}. The energy to extract a loop of length $\lambda_i$ is $\varepsilon_i$ which may be identified as a chemical potential. In order to obtain an expression for the total energy which considers the global features of the system,
we can claim that the mean energy to extract a single loop is $\varepsilon_0 =  E_0 / N$, where $E_0 = \sum_j \varepsilon_j$ is the total energy to pull out the whole structure. In a mean approach, the energy to extract $n$ loops is $E = n \varepsilon_0$, then
\begin{equation}
E = n \frac{E_0}{N} = \frac{E_0}{R_0} \ln \left( \frac{l+A}{A}\right),
\label{eq7}
\end{equation}
where we have used Eq.~\ref{eq4}. Thus, the first equality in Eq.~\ref{eq7} is a manifestation of the equipartition of energy among the loops. We stress that Eq.~\ref{eq7} is a better expression for the energy of this system when compared with Eq.~\ref{eq3}  because it takes into account: ({\it i}) the whole structure, ({\it ii}) the gradient of sizes of the loops, and therefore ({\it iii}) the confinement of the wire due to the cavity. The proportionality $E \sim n$ which supports Eq.~\ref{eq7} is also found for the elastic energy of the one-dimensional compactation of a thin sheet in $n$ layers. This process is indicated as a prototype for crumpling where the force is applied in one direction~\cite{Deboeuf13}.

The logarithm functional shape presented in Eq.~\ref{eq7} is not entirely new in the packing of wires. The elastic energetic cost of bending a straight wire into a spiral follows a similar relationship with the length of the wire~\cite{Bayart14}. Therefore Eq.~\ref{eq7} is in harmony with the idea that ordered models of folding are able to capture some features of crumpling~\cite{Deboeuf13}.

We can access some clues about the consecutive loops by adding the information obtained from Eq.~\ref{eq7} to the model where the loop is described by straight lines and circles as illustrated in Fig.~\ref{fig5}. In order to take in account the whole structure we extends Eq.~\ref{eq3} to an array of $N$ loops and a total arc length $L$.
The total bending energy over the wire can be taken loop-by-loop: 
\begin{equation}
E_b = I \int_{0}^L \kappa(l)^2dl = \sum_{i=1}^N I \int_{\{\lambda_i\}} \kappa_i^2 dl \sim \sum_{i=1}^N \kappa_i^2  \phi_ir_i,
\label{eq8}
\end{equation}
because the curvature is constant $\kappa_i = 1/r_i$ in the circular bulges of length $\phi_i r_i$. The energy of each loop is then $\varepsilon_i \sim \left( \phi_i / r_i \right)$. The mean field argument before Eq.~\ref{eq7} suggests that $\left( \phi_i / r_i \right)$ is roughly fixed and therefore the consecutive loops may have different elongations.

From the considerations of the previous paragraph, this is a conservative system and the extraction force $F$  can be taken as the negative of $- d E / d L$, where $L = L_n$ is the extracted length obtained by summing the size of the removed loops. From Eq.~\ref{eq7} follows
\begin{equation}
F = \left( \frac{d E}{d l} \right) \left( \frac{d l}{d L}  \right) \sim (l+A)^{-1}.
\label{eq9}
\end{equation}
The experimentally studied system involves an amount of energy dissipation through friction and plastic yielding that we believe is responsible for the shift from 1 to 0.9 found for the exponents in Eq.~\ref{eq6}. 

There was indicated in the literature some parallelism between the packing of crumpled wires in two dimensions and the three-dimensional DNA package in viral capsids~\cite{Maycon08,Katzav06}. Here we observe that the force dependence is similar in both cases. The elastic cost to bend a molecule of length $L$ into a circle of radius $R$ is $E_b \sim L / R^2$ immediately after Eq.~\ref{eq3}. Then the force is $F \sim R^{-2}$. Following the usual assumption in DNA models~\cite{Purohit05} this molecule twirls in the inner surface of a viral cavity in an ordered helicoidal configuration of radius $R$ and height $h$. The total number of hoops inside a capsule with constant shape is $N \sim h \sim R$ in such a way that the total length is $L=2\pi N R \sim R^2$. Therefore the total length is proportional to the area of the surface of the virus capsule as well as in the two-dimensional packing of wires. This reflects the dependence $F \sim L^{-1}$ for the DNA packaging in viral capsules and guide us in order to extend our results for crumpled systems in micro or even nano scales. 


\section*{Differential equation for the (un)packing of a rod}

The model used in this study is originally a discrete model: The loops are extracted from the cavity in steps of $\Delta n = 1$ with different sizes, $\lambda_n$, and the total length, $L_n$, is a geometrical series. In this sense the number of loops, $n$, is a discrete variable that goes until an unforeseeable number $N$. When we normalize, $n/N$, we still have a discrete variable, but when we consider a very large number of experiments this discrete behavior is diluted over a continuum of values. All fits made in our study were continuum ones, and they are in a very good agreement with the discrete expected behavior. In the thermodynamical regime, we expect $\Delta (n/N) \rightarrow 0$, while the variable $n/N$ runs continuously from zero to one. Combining Eq.~\ref{eq1} and Eq.~\ref{eq4} we can write
\begin{equation}
\lambda_n = \left( \frac{\lambda_0}{R_0 A} \right) l^{\prime}(n/N),
\label{eq10}
\end{equation}
where $l^{\prime}(n/N)$ denotes the derivative of $l$ with respect to $(n/N)$. Then, the function $l = l(n/N)$ obeys the differential equation
\begin{equation}
l^{\prime} = R_0 (l + A),
\label{eq11}
\end{equation}
where the exponential ratio $R_0$ is positive. The same equation can be written for the packing process, where $l$ is the fractional {\it packed} length and $n/N$ the fractional number of loops {\it inside} the cavity. In this case the exponential rate $R_0$ is negative because the loops progressively decrease, but the product $R_0 A$ remains positive. This term in Eq.~\ref{eq11} is important because the variation of the length have always a forced element due to the injection or extraction process.

Eq.~\ref{eq11} with $R_0<0$ allows us to make an analogy between the packing of wires in two-dimensional cavities and an RC circuit from basic physics. The charge in the capacitor corresponds to the total length $l$ in the cavity while the time corresponds to the number of loops $(n/N)$. The product $R_0 A$ is seen as due to a ``battery'' potential. For a given time $n$, this mass-capacitor has a charge $l$ and a current $l^\prime$ in such a way that the structure holds the wire inside the cavity as we can see from the static equilibrium shown in Fig.~\ref{apparatus}. If we proceed in our analogy we should expect that the energy $E$ is proportional to the square of the charge, here $E \sim l^2$. This last expression is in agreement with the self-exclusion energy commonly used to model this experiment~\cite{Maycon08,Gomes10,Gomes13}. We conjecture that the study of the packing of a wire in a two-dimensional cavity as a capacitor could be useful for the problem of delivery of biopolymers or polymers in biological tissues with the aid of natural or artificial nano carriers~\cite{DDS4,DDS5,DDS6,DDS7,DDS8,DDS9}.

\section*{Conclusions}

The study of crumpled structures of wires is a recent topic of wide interest from the points of view of theory and application \cite{Donato03,Stoop08,Gomes10,Gomes13,Cunha09}. Here it has been studied experimentally the progressive extraction of such a structure initially confined in two-dimensional cavities. The average profile of sizes of the constitutive loops was obtained and the force needed to unpack the loops was found for two different methods: as a function of the total length of the extracted wire as well as a function of the size of each loop. We have compared a model for a single loop with a mean-field argument for $n$ loops and concluded that our results for the dependence of the unpacking force is better explained by the collective behavior of loops. A hypothesis that each loop occupies a fixed fraction of the available space of the cavity discounting early loops lead us to an exponential model for the size of the loops and have yielded a differential equation for crumpled wires. Besides its intrinsic interest, the problem investigated here is of relevance for the unpacking of DNA from viral capsids \cite{Purohit05,Maycon08}, as well as for the important problem of delivery of polymers and one-dimensional bio-polymers into specific biological tissues by artificial or natural nano-structures  \cite{DDS1,DDS2,DDS3,DDS4,DDS5,DDS6,DDS7,DDS8,DDS9}. 

\section*{Acknowledgments}

We would like to acknowledge the anonymous reviewers whose questions and notes helped us to clarify critical arguments in several parts of this manuscript. We acknowledge Conselho Nacional de Desenvolvimento Cient\'ifico e Tecnol\'ogico (CNPq), Programa Nacional de Coopera\c{c}\~ao Acad\^emica (PROCAD), and Programa de N\'ucleos de Excel\^encia (PRONEX), all Brazilian Government Agencies, for financial support. 

%
%

\end{document}